# Magnetization and magnetoresistance in insulating phases of SrFeO$_{3-\delta}$

S. Srinath,[1] M. Mahesh Kumar,[2] M.L. Post,[2] and H. Srikanth[1,*]

[1]Materials Physics Laboratory, Department of Physics, University of South Florida,
Tampa, FL 33620, USA
[2]Institute for Chemical Process and Environmental Technology,
National Research Council of Canada, Ottawa, ON, K1A 0R6, Canada

## ABSTRACT

We report the synthesis and properties of two new insulating phases of SrFeO$_{3-\delta}$ with introduction of oxygen deficiencies in metallic SrFeO$_3$ ; one with $0.15 \leq \delta \leq 0.19$ (sample A) and the other above $\delta = 0.19$ (sample B). Sample A shows large negative magnetoresistance around the charged ordering (CO) temperature with magnetic anomalies seen in the temperature dependent resistivity, magnetization and M-H hysteresis loops. Sample B shows a smooth insulating behavior with no thermal hysteresis in the resistivity and with a small positive magnetoresistance. $\chi_{ac}$ and $\chi_{dc}$ show multiple features associated with a frustrated magnetic order (helical) due to competing ferro- and antiferromagnetic interactions. The competing effects of ferro- and antiferromagnetic phases extend up to T ~ 230 K revealing a new high temperature scale in this system. These observations are discussed in the context of magnetic interactions associated with the varying Fe$^{4+}$/Fe$^{3+}$ ratio.

[*] Corresponding Author: sharihar@cas.usf.edu

**Introduction**

Recent interest in materials which show colossal magnetoresistance (CMR) stems from their ability to produce large changes in resistance with the application of magnetic fields. These materials have also generally displayed a rich phase diagram with intricate coupling between electronic, magnetic and structural properties[1-5]. Among the very few materials that show coexistence of charge ordering and ferromagnetism with a metallic phase, $PrCa_{1-x}Mn_xO_{3-\delta}$ is notable. Compounds of Ruddlesden-Popper (RP) type, which are close to the manganites, fall into a category of materials with similar magnetic properties.[7] In particular, RP phases with $Fe^{4+}$ in a high spin state offer an interesting counterpart in ferrates.[8]

$SrFeO_3$ is a perovskite block in the RP class of systems in which the structural morphologies critically depend on the oxygen stoichiometry[9,10]. The charge ratio, $Fe^{4+}$ to $Fe^{3+}$ imparts a profound influence on the magnetic structure and originating properties in $SrFeO_{3-\delta}$. Stoichiometric $SrFeO_3$ ($\delta \sim 0$) is an antiferromagnet ($T_N \sim 140$ K) with a cubic perovskite structure at room temperature. An important feature of this ferrate is that it has a helical magnetic spin structure with a propagation vector parallel to the crystallographic [111] direction and shows no JT distortion even at very low temperatures.[7,11] $SrFeO_3$ however is metallic unlike its counterparts in manganites, with strong covalency having the $e_g^*$ orbitals extended into the itinerant conduction band with low electron densities around the nucleus.[12] Recently, it was shown by Lebon *et al*.,[13] that in certain oxygen deficient compositions of $SrFeO_{3-\delta}$, a charge ordered (CO) state with a giant negative magnetoresistance could be observed, which was attributed to the $Fe^{4+}$-$Fe^{3+}$ charge ordering.[13] The metal-insulator transition occurs around $\delta = 0.15$ with a magnetic transition around 70 K. With additional loss of oxygen ($\delta = 0.19$), an insulating behavior was seen with a large positive magnetoresistance at low temperatures and negative magnetoresistance near the magnetic transition (60 K). Two facts that are unclear in the above study are (a) the precise oxygen stoichiometry where the metallic SrFeO$_3$ transforms to a non-stoichiometric charge ordered insulator and (b) the nature of the intermediate composition bounded by stoichiometries that display metal-insulator transition ($\delta$= 0.15) and partly insulating behavior ($\delta$= 0.19). Our present investigation addresses these issues and attempts to isolate the intermediate phases (between $\delta$ = 0.15 and 0.19) and one above $\delta$ = 0.19 in SrFeO$_{3-\delta}$. Through systematic DC magnetization, AC susceptibility, zero-field resistivity and magnetoresistance measurements, we were able to identify the compositions in question and also found additional important magnetic transitions not reported earlier.

**Experimental**

Samples of two different oxygen stoichiometries of SrFeO$_3$ were prepared by solid-state reactions. Starting materials of SrCO$_3$, Fe$_2$O$_3$ were weighed in stoichiometric proportions, thoroughly mixed in an agate mortar and pre-fired at 1000 $^o$C. The powders were again mixed and pressed into pellets of 12 mm diameter and were fired at two different temperatures. One pellet was fired at 1300 $^o$C (Sample A) and the other was fired at a temperature of 1150 $^o$C (Sample B). Both were sintered in flowing oxygen in a tube furnace, with a reduced oxygen partial pressure for Sample B, in order to impart greater oxygen loss and enhance the Fe$^{4+}$/Fe$^{3+}$ ratio. Oxygen stoichiometry was estimated by reducing SrFeO$_3$ to its basic oxide and elements, through the associated weight losses, using the TA instruments' thermogravimetry analysis (TGA) with an accuracy of around 0.03. X-ray diffraction (XRD) was used to characterize the structure and phase purity of the materials was obtained using a Bruker D8 diffractometer. Magnetization, magnetic susceptibility (DC and AC), resistivity and magnetoresistance measurements were carried out in a Physical Property Measurement System (PPMS) from Quantum Design, as a function of temperature in the range, 10 – 300 K. Frequency was varied from 10 Hz – 10 kHz in the AC susceptibility measurements at fixed AC field amplitude of 10 Oe. DC magnetization was done in fields ranging from 0 to 6 T with the samples cooled in zero field at different temperatures. Resistivity as a function of temperature was measured in zero field and at 6 T using a standard four-probe technique.

**Results and discussion**

X-ray diffraction patterns obtained for Samples A and B (Inset, Fig. 1 (a)) showed a single $SrFeO_3$ phase without any traces of impurities. A comparison of the spectra of Samples A and B with the ideal cubic perovskite of $SrFeO_3$ indicates a slight shift of Bragg reflections to lower angles indicating elongation of the unit cell. An enlarged view of the (200) peak $q_B \sim 47^o$, shows that Sample A has a shoulder indicating symmetry lower than that of the cubic $SrFeO_3$. This signature of the pseudo-cubic phase continues through into Sample B which shows a clear splitting of the (200) reflection, indicating the advent of the tetragonal phase. Reports indicate the cubic ↔ tetragonal distortion to occur at $\delta = 0.15$ and a strong split in the reflections of Sample B indicate an oxygen stoichiometry either close to, or in excess of, $\delta = 0.15$. TGA showed a stoichiometry of $\delta = 0.17$ for Sample A and $\delta = 0.205$ for Sample B.

Figure 1 shows variation of AC magnetic susceptibility ($\chi'$) as a function of temperature for samples A and B measured at different frequencies. With lowering of temperature, Sample A (top panel) shows a transition around 78 K followed by a steep decrease in the magnetization down to 68 K. Further lowering of temperature results in a more gradual decrease in magnetization. Susceptibility curves at different frequencies ($0.5 \leq \nu \leq 10$ kHz) show no variation in the entire temperature range of measurement. In addition to the transition around 78 K, another transition could also be seen as a cusp at around 120 K. These two transitions can be identified as the T (tetragonal, partial $Fe^{3+}$ state) and a residual C (Cubic, $Fe^{4+}$ state) phase, respectively. In contrast, in an earlier report these two phases were found at ~82 K and ~ 130 K in $SrFeO_{2.85}$.[13]

Sample B, however, shows a clear cusp indicating the C phase around 110 K and a broad maximum (compared to Sample A) around 68 K indicating the T phase. The shape of this curve is close to that reported for $SrFeO_{2.81}$.[13] The transition around 110 K in Sample B is strong when compared to a slope change seen in $SrFeO_{2.81}$. These features indicate the oxygen stoichiometry of Sample B is close to but lower than $SrFeO_{2.81}$ in agreement with TGA data. Resistivity measurements also confirm that Sample A has a composition just above that of $\delta = 0.15$, and Sample B above that of $\delta = 0.19$.

Fig. 2 shows the resistivity and magnetoresistance curves measured at zero field and 6 T for Samples A and B. Resistivity increases as a function of temperature as Sample A (Fig.1 (a)) is cooled through its T phase transition temperature (~73 K) with a sudden increase at this temperature by at least an order of magnitude indicating the inter-relationship between the magnetization and transport anomalies. Resistivity increases exhibiting a power-law behavior on further cooling. On heating, the sample shows a hysteresis around the ~ 73 K transition. No such anomaly is seen at the 120 K transition indicating that this antiferromagnetic ordering is not mediated by conduction electrons. An important observation of this behavior when compared to the metal-insulator transition shown by $SrFeO_{2.85}$ in Ref.13, is that we do not see any low temperature metallic phase. This is consistent with a decrease in the Sample A's oxygen content and the $Fe^{4+}/Fe^{3+}$ ratio.

Application of a magnetic field of 6 T suppresses the resistance of Sample A by more than 60% around the 68 K transition (inset of Fig. 2, top panel). The negative magnetoresistance continues well down to low temperatures, except for a small region between 10 and 20 K where it becomes positive. In the report of Lebon *et al.*,[13] the low temperature magnetoresistance of $SrFeO_{2.85}$ shows no magnetoresistance at all whereas $SrFeO_{2.81}$ shows a positive magnetoresistance below 25 K.[13] It is interesting to note that the shape of the magnetization curve is close to that of reported $SrFeO_{2.85}$ and resistivity curve to that of $SrFeO_{2.81}$. These similarities along with values of $\delta$ obtained from TGA help to infer that the composition of Sample A falls between the reported $SrFeO_{2.85}$ and $SrFeO_{2.81}$.[13]

The resistivity of Sample B (Fig. 2, bottom panel) shows a monotonic insulating behavior from room to low temperatures. An interesting observation is that on the application of a 6 T magnetic field, the resistivity shows a positive magnetoresistance (see inset) of ~5% at 12 K. A fully insulating phase is only possible beyond the partly insulating $SrFeO_{2.81}$ phase and extending similar arguments to Sample B, it can be concluded that the composition of Sample B would have $\delta \geq 0.19$. The tetragonal structural distortion of Sample B (Fig. 1 inset) also supports this conclusion about the stoichiometry.

Figure 3 shows the zero field cooled (ZFC) and field cooled (FC) DC magnetization measurements in the temperature range 10-300 K. For sample A, ZFC data qualitatively shows all the features that were observed in the AC magnetic susceptibility (Fig. 1). The FC and ZFC curves are separate till about 230 K and merge above this temperature. The separation depends on the applied field and decreases as the field is increased to 1 T. The ZFC/FC curves of Sample B show a similar trend for different fields with the temperature at which the curves merge being 40 K for 100 Oe applied field.

The ZFC curve of 100 Oe for Sample A shows a distinct kink around ~230 K, in addition to the anomalies observed at lower temperature. This is also seen in the ZFC as well as FC measurements at other fields (Fig 3). This new feature is completely unexpected, in view of the similar measurements carried out by many authors on this system.[7,12,13] In order to explore this further, the magnetic hysteresis measured at different temperatures spanning a range from 10K to 300K are shown in Fig. 4 for Sample A. At low temperatures (T~10 K) an extremely small loop could be seen, which does not saturate even at fields of 6 T. This is due to the presence of ferromagnetic interactions in the system. Note that there is a strong shift of the M-H curves away from the origin as the temperature is systematically lowered below 230 K. This tendency is associated with the coexistence of antiferromagnetic and ferromagnetic interactions and is typical in materials with helical magnetic ordering. For example, such loops are seen in MnSi which is a classic example of a helical magnet.[14] Competing ferro- and antiferromagnetic effects due to the helical spin structure gives rise to a frustrated system similar to a spin glass and an unsaturated magnetization is caused by the spin fluctuations in the helical spin arrangement.

While the shift in the M-H loops due to exchange bias effects are expected below $T_N$ which is 120K for Sample A, the fact that a systematic shift of the loops is observed even up to T = 230 K, indicates that the coexistence of competing magnetic order persists up to this temperature. It is possible that due to the presence of this magnetic ordering at relatively high temperatures, attempts by several authors to fit the susceptibility data above the Neel temperature to Curie-Weiss law only resulted in a curvature, without an acceptable degree of fit. The inverse susceptibility ($\chi^{-1}$) data above 230 K can be described by using the Curie Weiss law $\chi = C/(T-\theta)$, and the fit to the data is shown in Fig. 3 inset. The parameters obtained from the fit are C = 2.47(1) and $\theta$ = 20.8 (1) K. An effective moment, $\mu_{eff}$, of 4.44 $\mu_B$, is obtained from the equation $\mu_{eff} = \sqrt{7.99C}$. This is somewhat smaller than 4.9 $\mu_B$ which is the spin-only value for $Fe^{4+}$. In the case of Sample B, similar analysis yielded a larger effective moment (4.9 as opposed to 4.44 for Sample A) indicating the increased contribution due to $Fe^{3+}$.

We can now put all these results in perspective and try to develop an understanding of magnetism and transport in these materials. It is known that the magnetic anomalies seen in this system are mainly due to the presence of varying proportions of $Fe^{4+}/Fe^{3+}$ content in oxygen deficient samples.Prior reports[12] also indicate the existence of fractional valence ($Fe^{3+\Delta}$) as a function of temperature that would profoundly affect the magnetic properties. This valence fluctuation adds to the complexity of a complete understanding of the strong correlation between electronic spin and charge leading to exotic magnetic and transport properties.

The sharp rise in resistivity of Sample A, thermal hysteresis and associated magnetic anomalies are reminiscent of charge ordering (CO). In that sense, these systems have similarities to the perovskite oxide systems such as $Pr_{1-x}Ca_xMnO_3$ and $La_{1-x}Sr_{1+x}MnO_4$ which are known to exhibit charge ordering, large negative magnetoresistance etc.[15,16]

In Sample B, the oxygen deficiency is larger that that for Sample A which reduces the overlap of Fe-Fe orbitals even more. At the same time, the $Fe^{4+}/Fe^{3+}$ ratio changes because of increase in $Fe^{3+}$ which contributes to excess electrons in the $e_g$ band. This combined effect results in insulating behavior, albeit with a lower resistivity. No CO seems to occur in this sample with a logarithmic increase in resistivity with decreasing temperature. The resistivity data between 76 and 300 K was found to fit to the Variable Range Hopping (VRH) model.

The irreversibility in ZFC and FC curves in Sample A indicate the magnetic frustration induced by the competing affects of inter layer ferromagnetism (FM) and intra layer antiferromagnetic (AFM) order due to the helical magnetism. The degree of irreversibility is high for low applied fields and persists up to high temperatures as seen in Figure 3a. With increasing magnetic field, the degree of irreversibility decreases. This is consistent with the helical magnetic order that would transform to a collinear order at high fields. In $SrFeO_{3-\delta}$ a spin angle between neighboring {111} planes has been found to be $q = 46°$. Strong magnetic fields can decrease the $q$, reducing the irreversibility.[17]

In comparison, Sample B exhibits a smaller degree of irreversibility, which occurs only at low temperatures, indicating the reduced frustration and possible melting of the helical magnetic spin structure. This dilution of the helical spin structure is possible with increasing content of $Fe^{3+}$ and its interaction with $Fe^{4+}$ in $SrFeO_{3-\delta}$, which could yield an overall antiferromagnetic ordering.

We believe that the origin of the 230 K transition in Sample A and B result from the exchange interactions that dominate at temperatures between 70 and 230 K. The interactions are between $Fe^{4+}$-O-$Fe^{4+}$ and $Fe^{4+}$- O - $Fe^{3+\Delta}$, which are antiferromagnetic (former) and ferromagnetic (latter) in nature.[18] The positive intercept of the Curie-Weiss fit, hysteresis in the M-H curves and the shift of the overall M-H loops along the field axis observed in our data up to 230 K are all consequences of competing ferro and antiferromagnetic effects.

**Acknowledgements**

Work at USF is supported by DARPA through ARO Grant No. DAAD19-03-1-0277. Authors wish to thank Mr. Kai Herberz for assistance in sample preparation and Dr. Serguei Koutcheiko, ICPET, Ottawa, Canada for the help in TGA measurements. The financial support of this project by National Research Council of Canada, in the NRC-Helmholtz collaborative program is gratefully acknowledged (NRCC-21-CRP-02).

**Figure Captions**

FIG.1  Ac magnetic susceptibility as a function of temperature for (a) Sample A and (b) Sample B. Inset (a): X-ray diffraction spectrum of the (200) plane for Sample A & B.

FIG. 2 Resistivity and magnetoresistance zero field and 6 T for (a) Sample A and (b) Sample B as a function of temperature. Insets show the percentage change in resistance on the application of 6 T magnetic field.

FIG. 3  Zero field cooled (ZFC) and field cooled (FC) measurements for Sample A at 1 Tesla, 1 kOe and 100 Oe, and, for sample B at 1 kOe and 100 Oe. A new transition observed T ~ 230 K is shown with arrows. For 1 Tesla, a hysteresis around this temperature (marked with an arrow) can be seen.

FIG. 4  Magnetization isotherms measured at different temperatures for Sample A. Only low field data is shown for clarity even though the data was acquired up to 6 T.

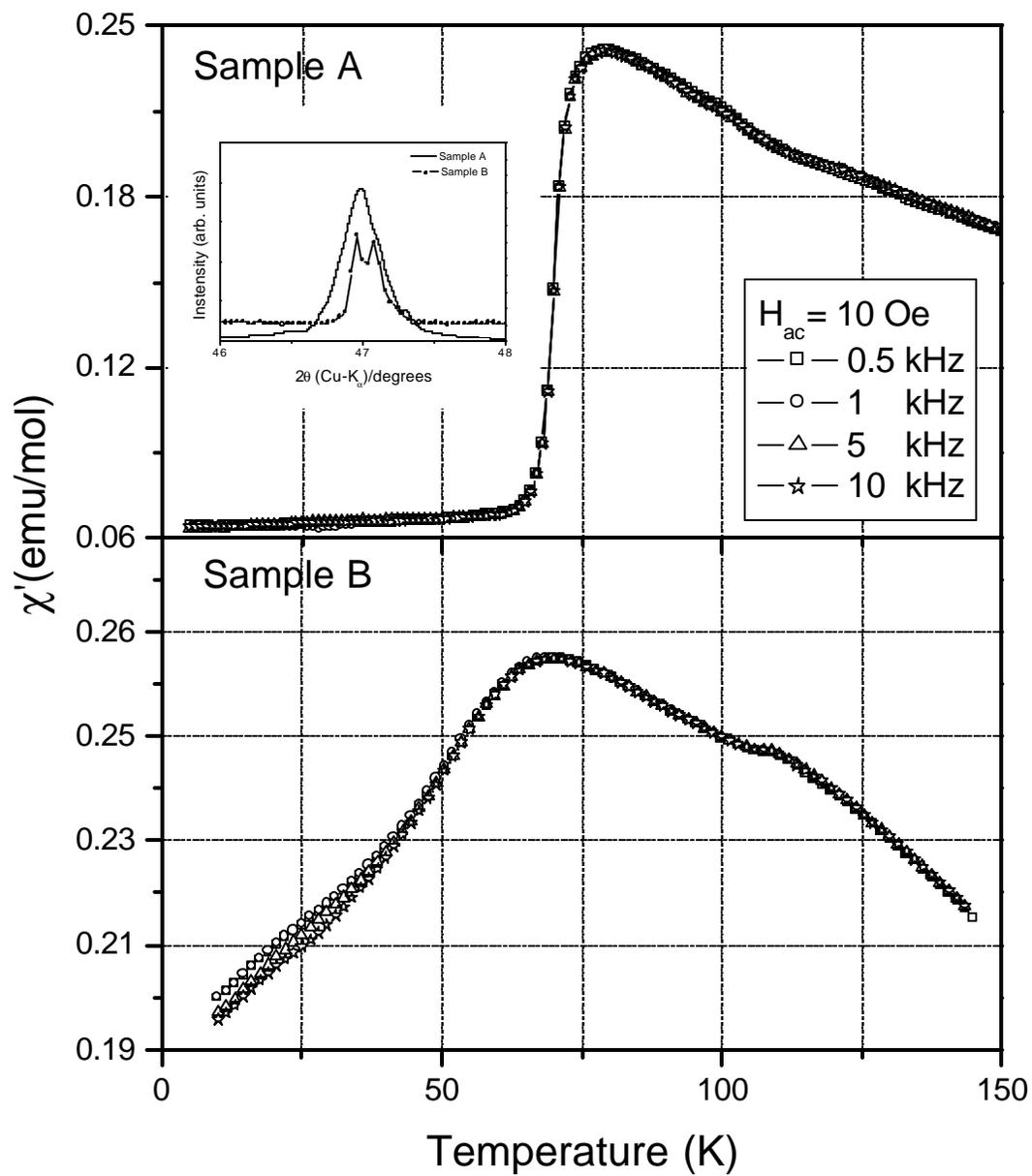

Figure 1

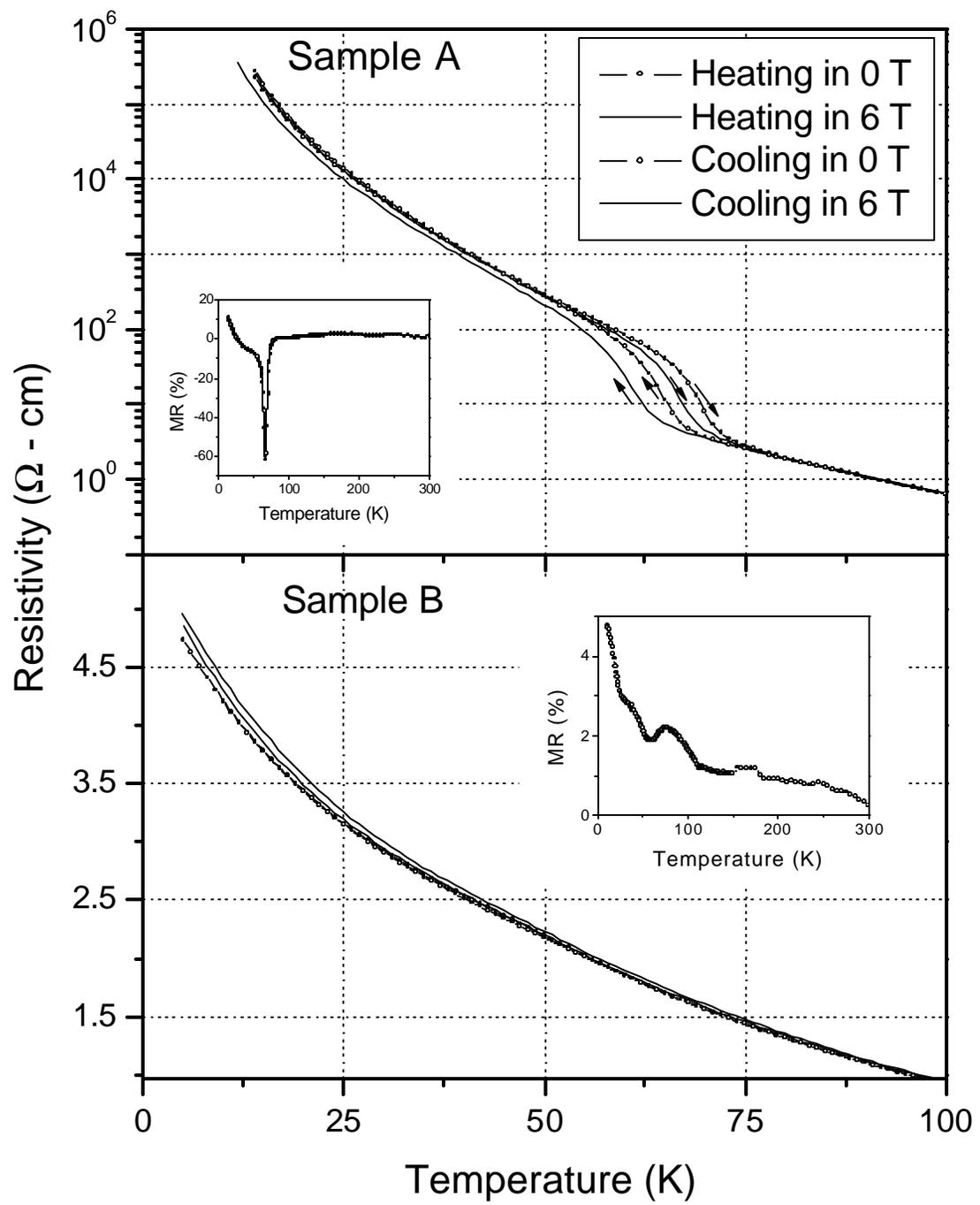

Figure 2

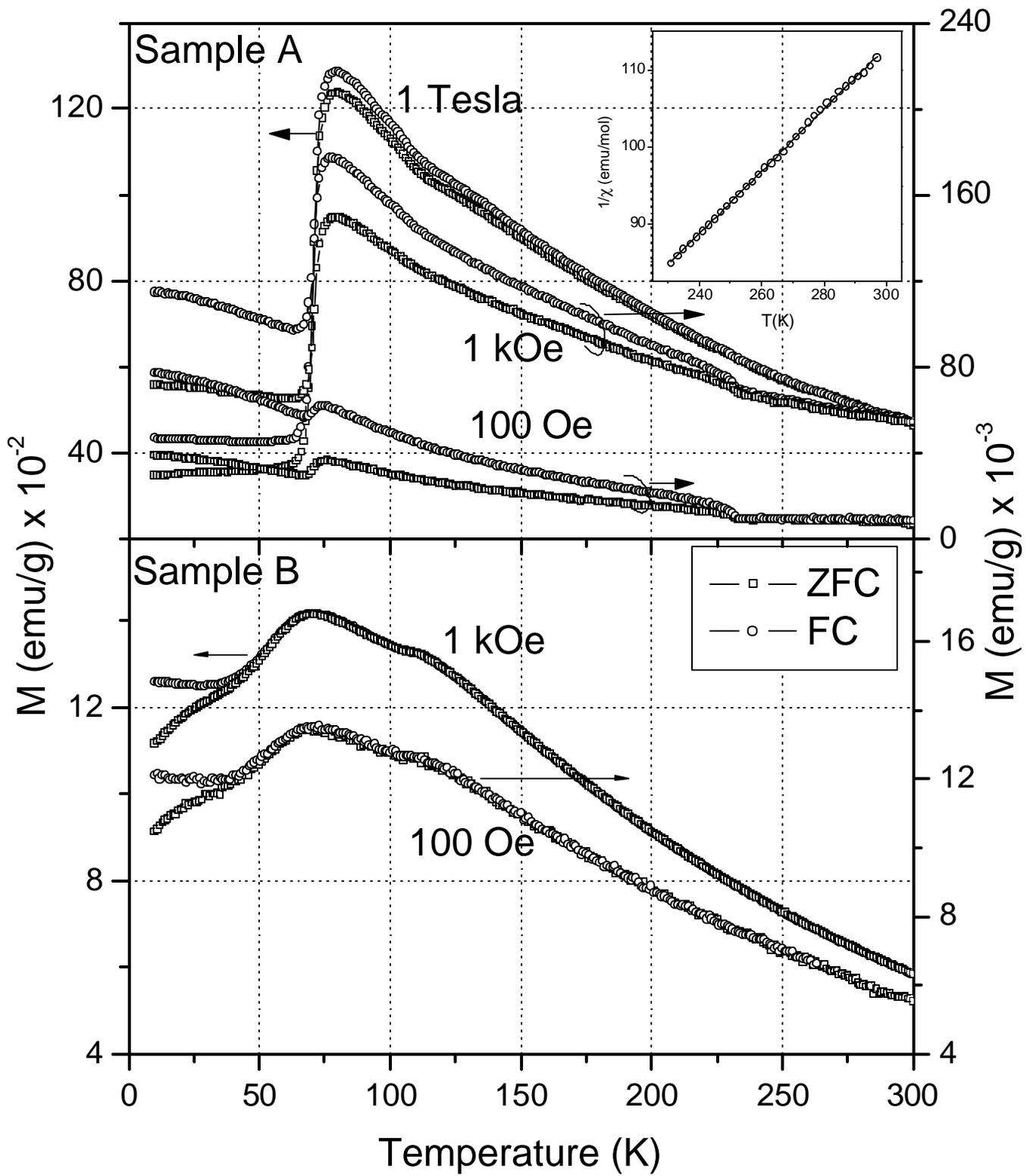

Figure 3

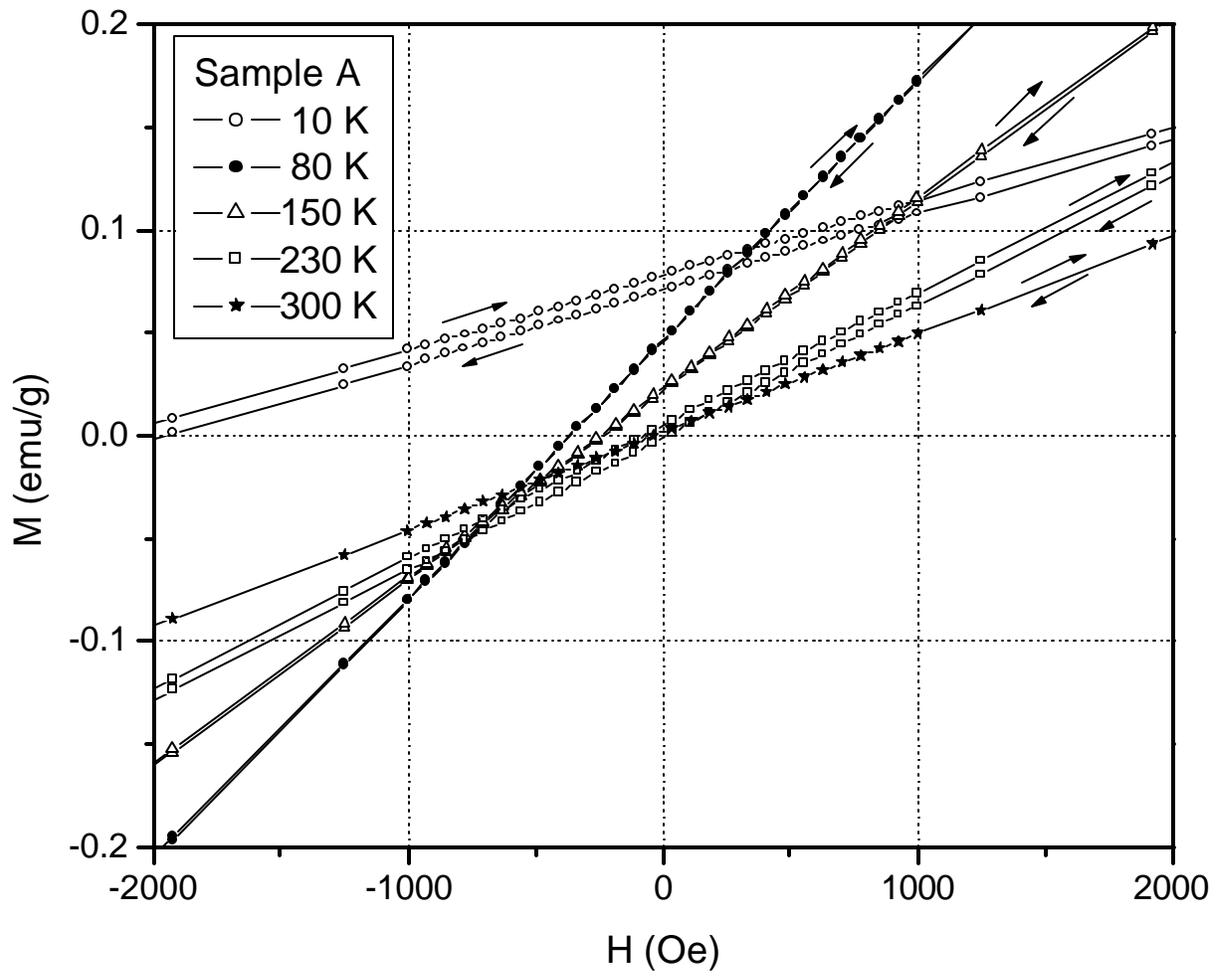

Figure 4